\documentclass[prl,epsf,showpacs,twocolumn,superscriptaddress,]{revtex4-1}
\usepackage[pdftex]{graphicx}
\usepackage{dcolumn}
\usepackage{bm}
\usepackage{epsfig}
\usepackage{latexsym}
\usepackage{amsmath}
\usepackage{amsfonts}
\usepackage{color}
\usepackage{array}
\usepackage{braket}
\usepackage{bbold}
\usepackage{dsfont}
\usepackage{mathrsfs}
\usepackage{hyperref}
\hypersetup{
    colorlinks,
    citecolor=blue,
    filecolor=red,
    linkcolor=magenta,
    urlcolor=red
}
\usepackage[dvipsnames]{xcolor}

\makeatletter
\DeclareFontFamily{OMX}{MnSymbolE}{}
\DeclareSymbolFont{MnLargeSymbols}{OMX}{MnSymbolE}{m}{n}
\SetSymbolFont{MnLargeSymbols}{bold}{OMX}{MnSymbolE}{b}{n}
\DeclareFontShape{OMX}{MnSymbolE}{m}{n}{
    <-6>  MnSymbolE5
   <6-7>  MnSymbolE6
   <7-8>  MnSymbolE7
   <8-9>  MnSymbolE8
   <9-10> MnSymbolE9
  <10-12> MnSymbolE10
  <12->   MnSymbolE12
}{}
\DeclareFontShape{OMX}{MnSymbolE}{b}{n}{
    <-6>  MnSymbolE-Bold5
   <6-7>  MnSymbolE-Bold6
   <7-8>  MnSymbolE-Bold7
   <8-9>  MnSymbolE-Bold8
   <9-10> MnSymbolE-Bold9
  <10-12> MnSymbolE-Bold10
  <12->   MnSymbolE-Bold12
}{}

\let\llangle\@undefined
\let\rrangle\@undefined
\DeclareMathDelimiter{\llangle}{\mathopen}%
                     {MnLargeSymbols}{'164}{MnLargeSymbols}{'164}
\DeclareMathDelimiter{\rrangle}{\mathclose}%
                     {MnLargeSymbols}{'171}{MnLargeSymbols}{'171}

\newcommand{\beginsupplement}{%
        \setcounter{table}{0}
        \renewcommand{\thetable}{S\arabic{table}}%
        \setcounter{equation}{0}
        \renewcommand{\theequation}{S\arabic{equation}}%
        \setcounter{figure}{0}
        \renewcommand{\thefigure}{S\arabic{figure}}%
     }

\newcommand{\fref}[1]{Fig.~\ref{#1}}
\newcommand{\eref}[1]{Eq.\,(\ref{#1})}

\newcommand{\Z}{\mathbb{Z}}

\begin{document}

\title{Entanglement from tensor networks on a trapped-ion QCCD quantum computer}

\author{Michael Foss-Feig}
\email{michael.feig@honeywell.com}
\author{Stephen Ragole}
\affiliation{Honeywell Quantum Solutions, 303 S. Technology Ct., Broomfield, Colorado 80021, USA}

\author{Andrew Potter}
\affiliation{Department of Physics, University of Texas at Austin, Austin, TX 78712, USA}

\author{Joan Dreiling}
\author{Caroline Figgatt}
\author{John Gaebler}
\author{Alex Hall}
\author{Steven Moses}
\author{Juan Pino}
\author{Ben Spaun}
\author{Brian Neyenhuis}
\author{David Hayes}
\affiliation{Honeywell Quantum Solutions, 303 S. Technology Ct., Broomfield, Colorado 80021, USA}

\begin{abstract}
The ability to selectively measure, initialize, and reuse qubits during a quantum circuit enables a mapping of the spatial structure of certain tensor-network states onto the dynamics of quantum circuits, thereby achieving dramatic resource savings when using a quantum computer to simulate many-body systems with limited entanglement.  We experimentally demonstrate a significant benefit of this approach to quantum simulation: In addition to all correlation functions, the \emph{entanglement structure} of an infinite system---specifically the half-chain entanglement spectrum--is conveniently encoded within a small register of ``bond qubits'' and can be extracted with relative ease. Using a trapped-ion QCCD quantum computer equipped with selective mid-circuit measurement and reset,  we quantitatively determine the near-critical entanglement entropy of a correlated spin chain directly in the thermodynamic limit and show that its phase transition becomes quickly resolved upon expanding the bond-qubit register.
\end{abstract}
\maketitle

Of the many applications considered for near-term quantum computers, the simulation of strongly correlated quantum systems stands out for being useful, hard classically,  and tolerant of at least some imperfections (nature is, after all, a noisy place).  Yet even in simulating quantum systems, a problem so well tailored to quantum computing that it is often credited with initiating the entire field, solving classically hard problems of real utility remains stubbornly out of reach. Part of the difficulty is that---unlike the outputs of random unitaries \cite{Arute:2019aa,Zhong1460}---states of physical quantum systems are highly structured; the best classical algorithms exploit this structure to a remarkable degree, setting a high bar for quantum advantage.  It seems likely that any near-term quantum advantage will strongly benefit from, and possibly even require, quantum algorithms that can also exploit this structure. Over the last several years notable progress has been made along these lines, with quantum algorithms designed around matrix-product states (MPS) \cite{Kim_2017a,Huggins_2019,Liu_2019,foss2020holographic,smith2019crossing,PRXQuantum.2.010342,barratt2020parallel}, tree tensor networks \cite{Huggins_2019,yuan2020quantum}, and variations on the multi-scale entanglement renormalization ansatz (MERA) \cite{Kim_2017}.  Here we demonstrate a remarkable feature of these algorithms: In addition to inheriting considerable resource savings from their classical precursors, they also provide a remarkably direct encoding of the entanglement structure in states they represent.
%
%
\begin{figure}[t!]
\begin{center}
\includegraphics[width=1.0\columnwidth]{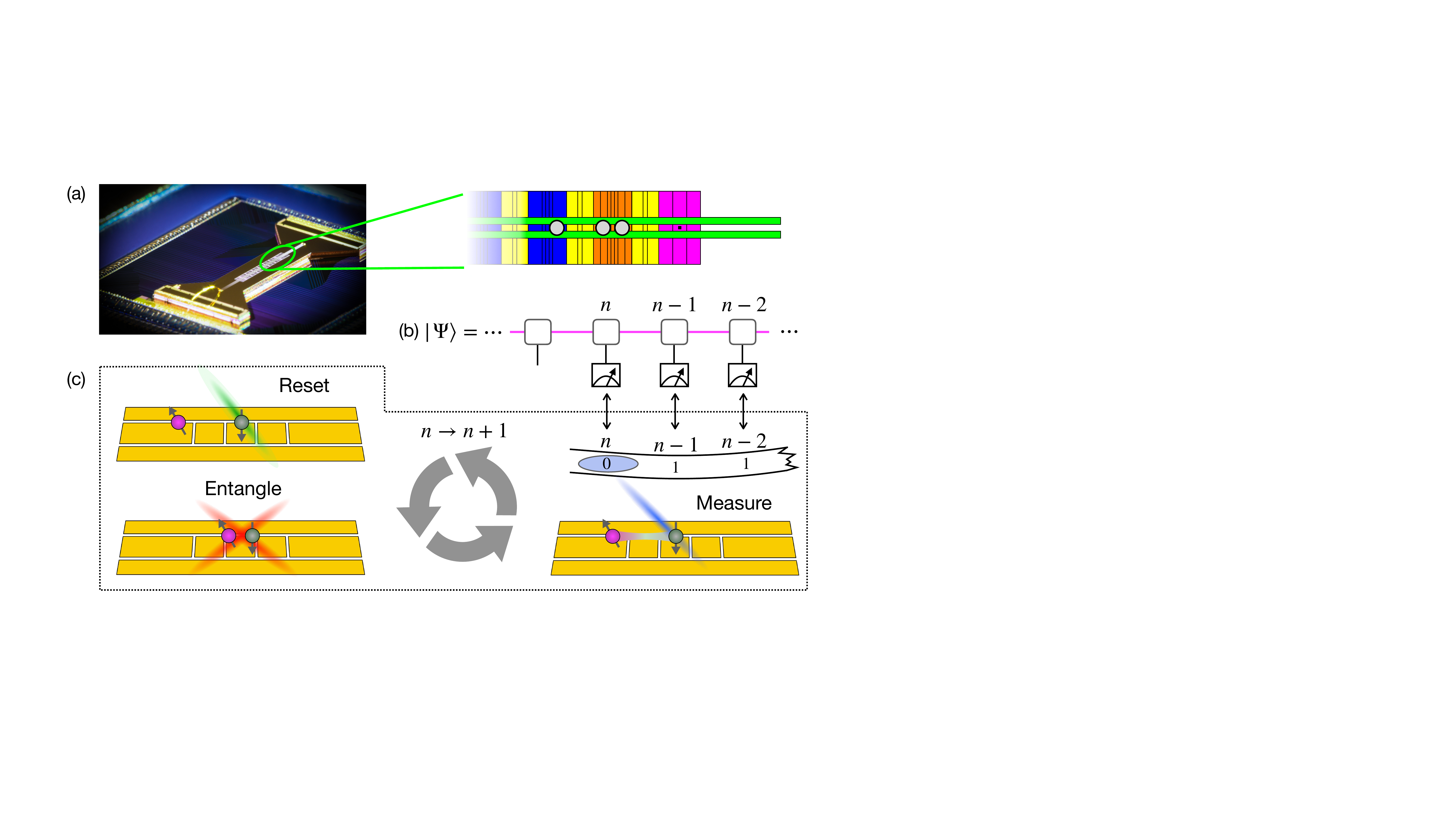}
\caption{Using Honeywell's trapped-ion quantum computer described in Ref. \cite{Pino:2021tf} and shown in (a), we execute a state-preparation algorithm for arbitrary matrix-product states (MPS) (b).  (c) The algorithm consists of repeatedly entangling a \emph{single} system qubit (grey) with a register of $\log\chi$ ``bond'' qubits (purple) in order to encode a bond-dimension $\chi$ MPS (here we show the case of $\chi=2$, which requires a single bond qubit).  The spatial extent of the MPS is encoded in the temporal extent of the algorithm, with properties of the $n^{\rm th}$ site corresponding to the measurement record of the system qubit prior to reset at the $n^{\rm th}$ iteration.
}
\label{fig:summary}
\end{center}
\end{figure}

The latter point is especially enticing, as many-body entanglement entropy offers valuable information-theoretic insights into the structure of complex quantum matter that cannot be captured by local correlations~\cite{kitaev2006topological,levin2006detecting}; understanding its structure sheds light on the quantum foundations of thermodynamic entropy~\cite{d2016quantum}, thermalization and quantum chaos~\cite{abanin2019colloquium}, and even perhaps the geometry of space-time itself~\cite{nishioka2018entanglement}. Moreover, universal scaling features of entanglement \cite{calabrese2004entanglement}, such as central charge and its higher dimensional analogs, have entanglement-based interpretations and serve as fingerprints of critical phenomena.

A key technical challenge in running many such algorithms is the necessity to perform selective mid-circuit measurement and reset/reuse (MCMR) of qubits during a quantum circuit.  Long recognized as a crucial ingredient for scalable quantum computation, this ability and other closely-related capabilities have been realized in several quantum computing platforms \cite{Barrett:2004vw,Pfaff:2013ue,Riste:2013um}. Trapped ions in particular afford several convenient and high-fidelity approaches, including dynamic spatial isolation \cite{Chiaverini:2004uj,Barrett:2004vw,Wan875}, shelving \cite{Monz1068}, and dual-species quantum logic gates \cite{Schmidt749,Rosenband1808,Negnevitsky:2018ur}. Until very recently \cite{Pino:2021tf,corcoles2021exploiting,chen2021exponential}, however, MCMR had not been implemented on commercial quantum computing hardware.

In this paper, we use a trapped-ion quantum computer equipped with MCMR [\fref{fig:summary}a] to implement an efficient quantum algorithm for extracting the near-critical entanglement entropy of correlated spin chains.  We apply this algorithm to the transverse-field Ising model (TFIM) because it is simple and has well-understood critical properties; however, the algorithm we describe is not restricted to integrable models, nor is it restricted to 1D \cite{foss2020holographic}.  In addition to extracting accurate estimates of the TFIM ground-state energy with remarkably few qubits, we are able to quantitatively observe the divergence of bipartite entanglement entropy upon approaching the quantum phase transition separating its ordered and disordered phases.  

In its simplest incarnation, the algorithm we employ involves one ``system qubit'', one ``bond qubit'', and repeated applications of the process shown in \fref{fig:summary}(c): (1) Reset the system qubit, (2) Apply a unitary entangling operation between the system qubit and bond-qubit, and (3) Measure the system qubit.  The bond register propagates spatial correlations of the MPS through time, and each successive (in time) measurement of the system qubit extracts local information about the next adjacent site in the spatial extent of the 1D TFIM, from which we reconstruct an experimental estimate of the ground-state energy.  Moreover, the reduced density matrix of the bond-qubit after the $j^{\rm th}$ iteration encodes the reduced density matrix of the half-chain containing sites up to and including $j$.  The number of bond qubits $n_{\rm b}$ determines the accessible MPS bond-dimension $\chi$, which grows exponentially as $\chi= 2^{n_{\rm b}}$, enabling an extremely rapid convergence of results in the number of available bond qubits. Our data show that even for $n_{\rm b}=2$ the divergence of entanglement entropy at the phase transition is quite sharply resolvable.

\emph{Quantum MPS}---Matrix product states are an ansatz designed to efficiently capture the properties of 1D systems with limited entanglement, and have been employed extensively in classical simulations of 1D and quasi-2D quantum systems \cite{schollwock2011density}. While many aspects of the discussion below generalize to more expressive tensor networks, including tree-tensor networks, MERA, and 2D isometric tensor networks, the key ideas are already contained in the simplest case of MPS, which are the focus of this demonstration.  An MPS $\ket{\Psi}$ of a half-infinite system can be written

\begin{figure}[t]
\begin{center}
\includegraphics[width=1.0\columnwidth]{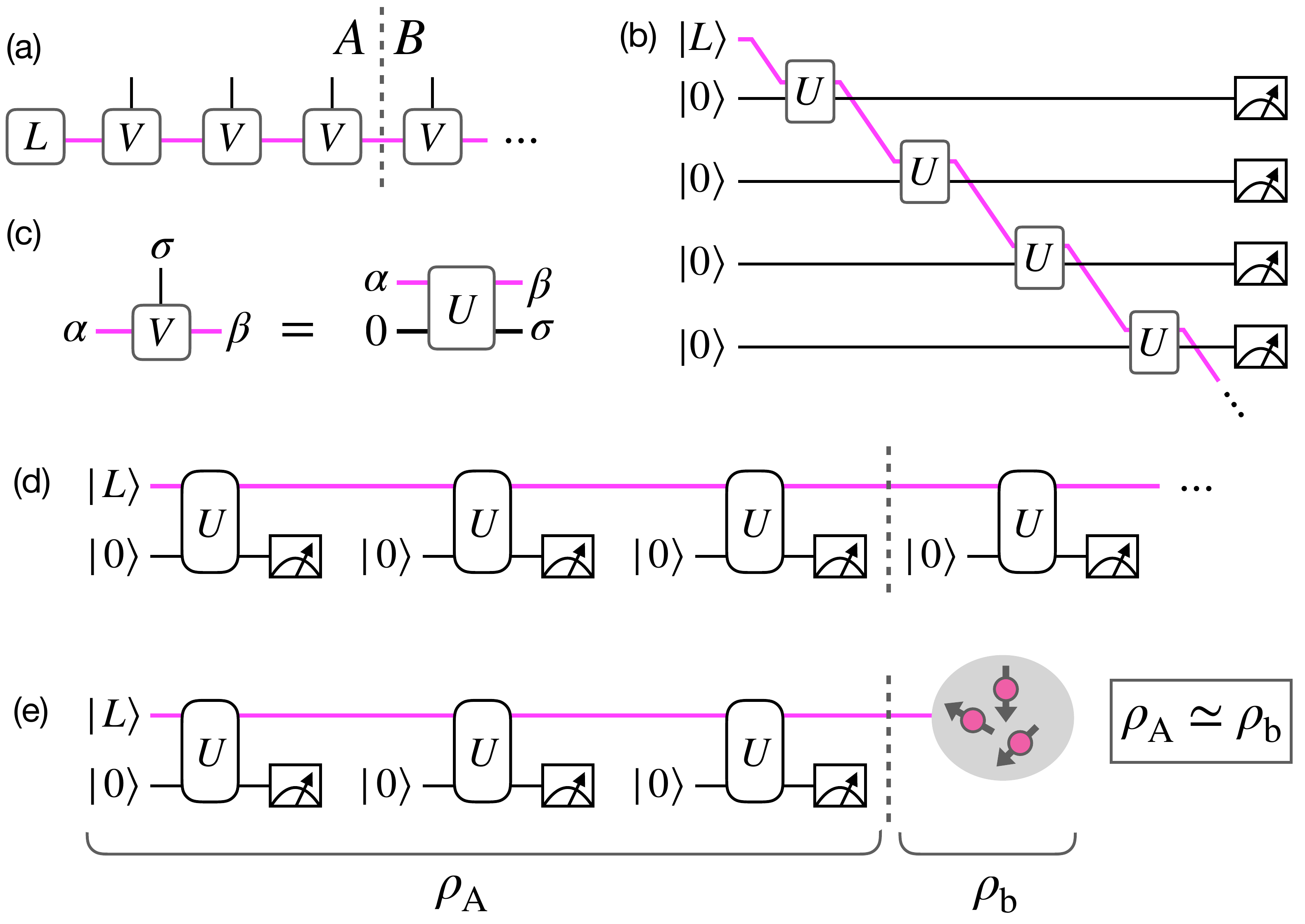}
\caption{Quantum MPS: A matrix product state (a) can be represented as a circuit (b) by embedding the tensors inside unitary gates according to \eref{eq:unitary_embedding} [represented graphically in (c) above]. By executing any desired measurements as soon as possible, the system qubit representing site $n$ can be reset and reused as the system qubit on site $n+1$, leading to an equivalent circuit with just a single system qubit (d).  Entanglement entropy of the MPS across a bipartition (dashed vertical line in the figures) becomes entanglement entropy of the bond qubits immediately after crossing that partition (e), as future gates cannot affect the state prior to the partition.}
\label{fig:mps_circuit}
\end{center}
\end{figure}
\begin{align}
\ket{\Psi}=\sum_{\sigma_1,\sigma_{2},\dots}L^{T}(V^{[1]}_{\sigma_{1}}V^{[2]}_{\sigma_{2}}\cdots )\ket{\sigma_1,\sigma_{2},\dots}.
\label{eq:general_mps}
\end{align}
Here, $\sigma_j$ indexes a set of basis states on site $j$, the $V_{\sigma_j}^{[j]}$ are matrices (matrix indices suppressed), and $L$ is a vector determining the left-boundary conditions. We will consider MPS with spatially uniform tensors and thus drop the site-dependent superscripts; sufficiently far from the left boundary, such MPS will have discrete translational invariance and thus encompass an important class of states for simulating naturally occurring systems.  We also restrict our attention to qubits as the physical degrees of freedom, in which case the physical indices take values $\sigma_j=0,1$ \footnote{Systems with larger local dimension can also be simulated using qubits by associating more than one physical qubit to a lattice site}.  The quantity $V_{\sigma}$ is then a set of 2 matrices [with matrix multiplication implied in \eref{eq:general_mps}], or alternatively (restoring matrix indices as superscripts) a rank-3 tensor $V_{\sigma}^{\alpha\beta}$. The MPS can be drawn schematically as in \fref{fig:mps_circuit}(a), where each tensor is a box with a leg for each index, and joined legs imply tensor contraction of the associated indices.  Without loss of generality the tensor $V$ can be chosen to be an isometry \cite{schollwock2011density}, and can therefore be embedded in a unitary $U$ as [see \fref{fig:mps_circuit}(c)]
\begin{align}
\label{eq:unitary_embedding}
V_{\sigma}^{\alpha\beta}=(\bra{\sigma}\otimes\bra{\beta})U(\ket{0}\otimes\ket{\alpha}).
\end{align}
Using this embedding we can construct the MPS of \eref{eq:general_mps} with the following quantum circuit involving a bond register interacting sequentially with the system qubits one by one,
\begin{align}
\ket{\Psi}=\cdots U_{2,{\rm b}}U_{1,{\rm b}}(\cdots\otimes\ket{0}_2\otimes\ket{0}_{1})\otimes \ket{L}_{\rm b},
\end{align}
as shown in \fref{fig:mps_circuit}(c).  Note that the first (top-most) system qubit is immediately gated with the bond-qubit register and then is idle for the rest of the circuit.  We are therefore free to measure it before the second qubit is initialized and gated with the bond register, so we might as well reinitialize the first qubit after measurement and then relabel it as the second qubit, \emph{as long as we can perform the measurement and reset in a way that does not perturb the bond qubits}. Iterating this logic, arbitrary local measurements on $\ket{\Psi}$ prepared by the circuit in \fref{fig:mps_circuit}c can be made using just one system qubit via the circuit in \fref{fig:mps_circuit}d.



Any MPS in canonical form is equivalent to a quantum channel defined on the bond indices, which get reinterpreted as labeling states in a fictitious bond Hilbert space \cite{schuch2011classifying}. In the method just described for generating an MPS with one system qubit, the process of initializing the system qubit, gating it with the bond qubit register, and then ignoring it (tracing it out) is precisely the unitary embedding (Stinespring dilation) of this bond-space quantum channel.  A convenient consequence  of this equivalence is that the bipartite entanglement spectrum induced by cutting an infinite chain into two halves $A$ and $B$, i.e., the eigenvalues of $\rho_A={\rm Tr}_{B} (\ket{\Psi}\!\bra{\Psi})$, can be extracted from the steady state of this quantum channel \cite{Gopalakrishnan_2019}.  The relationship between the entanglement spectrum of an MPS and the steady state of its associated channel is readily apparent in the unitary embedding employed here. The infinite MPS case is well approximated by the half-infinite MPS we generate if we place the cut sufficiently far from the left boundary and then trace out the right (or left) half, as in \fref{fig:mps_circuit}a.  Figure \ref{fig:mps_circuit}d shows the state preparation scheme used here with the equivalent cut.  Because the left half of the chain is fully formed immediately after the bond qubits pass the cut, its entanglement entropy can only result from entanglement between it and the bond qubits, and can be extracted at that point without ever building the right half of the state.  By the symmetry of entanglement spectra across the cut, the spectra of $\rho_{A}$ is then identical to that of the bond-qubit register at this point, as shown in \fref{fig:mps_circuit}e.


\emph{Demonstration}.---As an example of the MPS preparation technique and entanglement entropy extraction, we use a QCCD trapped-ion quantum computer \cite{Pino:2021tf} to construct MPS approximations to the ground-state of the transverse-field Ising model (TFIM), with Hamiltonian
\begin{align}
H=-\sum_{j}(Z_{j}Z_{j+1}+\lambda X_j).
\end{align}
For transverse-field strengths $\lambda>1$ the TFIM ground state is disordered with respect to the $\Z_2$ symmetry of the Hamiltonian, while for $\lambda<1$ the ground state spontaneously breaks that symmetry, and these phases are separated by a continuous quantum phase transition. When approaching the transition from either side, the entanglement entropy diverges logarithmically \cite{calabrese2006entanglement},
reflecting the divergent correlation length at the transition and serving as a sensitive probe of the phase transition \cite{PhysRevB.78.024410}.

The quantum computer used in this work operates with up to 6 qubits and two gate zones.  Each qubit is encoded in the states $\ket{1(0)}\equiv\ket{F=1(0),m_F=0}$ of the $S_{1/2}$ ground-state hyperfine manifold of a $^{171}{\rm Yb}^{+}$ ion, with $F$ and $m_F$ quantum numbers of the total internal angular momentum and its projection along a $\approx 5\,{\rm G}$ applied magnetic field, respectively.  Qubit ions are co-trapped with an equal number of $^{138}{\rm Ba}^{+}$ ions, which are used to sympathetically cool the motion of the qubits during a quantum circuit without impacting their logical state.  All laser-based operations (gates, measurement, and reset) are carried out in the two gate zones shaded blue and orange in \fref{fig:summary}a.  The ions are stored in either Ba-Yb or Ba-Yb-Yb-Ba crystals, with single-qubit gates (average fidelity $\approx 99.97\%$) performed on the former and two-qubit gates (average fidelity $\approx 99.2\%$) on the latter.  Arbitrary connectivity of two-qubit gates is achieved by physically rearranging the qubits between the various crystals into suitable pairs prior to gating the pairs within each crystal.

For the present demonstration, it is crucial that a subset of the qubits (in this case the system qubit) can be selectively measured and reset in the middle of a circuit without impacting the remaining qubits, which effectively encode the MPS and its entanglement structure.  Both reset and measurement involve applying light to resonantly excite the $\ket{1}$ state outside of the qubit subspace, either on a cycling transition (measurement) or to optically pump $\ket{1}\rightarrow \ket{0}$ state \cite{PhysRevA.76.052314}.  Even a single resonant photon scattered by a bond qubit amounts to an error on that qubit; fortunately the QCCD architecture affords multiple tools to suppress such crosstalk.  Most importantly, qubits that are measured or reset in the middle of a circuit are temporarily isolated \cite{Chiaverini:2004uj,Barrett:2004vw,Wan875} from all other qubits by at least $110\,\mu {\rm m}$ during the reset process, whereas the reset beams have a $1/e^2$ radius of $\approx 18\,\mu {\rm m}$ \footnote{The measurement and reset beams propagate at a $45^{\circ}$ angle to the inter-crystal axis, and therefore the effective $1/e^2$ radius determining optical crosstalk between crystals is larger than the quoted value by a factor of $\sqrt{2}$}.  MCMR crosstalk is further suppressed by using independent electrode control to push unmeasured qubits off the trap axis, inducing micromotion due to the RF trapping potential.  This motion effectively causes the measurement/reset laser to appear phase modulated in the frame of the unmeasured qubits, displacing a large fraction of the already low laser intensity into off-resonant sidebands.  Detailed analysis of the measurement and reset  cross talk in this system will be published elsewhere \footnote{Manuscript in preparation.}, but for our purposes it suffices to know that the average infidelity induced on spectator qubits due to reset(measurement) crosstalk is $\lesssim 4\times10^{-4}(2\times 10^{-3})$.

\begin{figure}[t]
\begin{center}
\includegraphics[width=1.0\columnwidth]{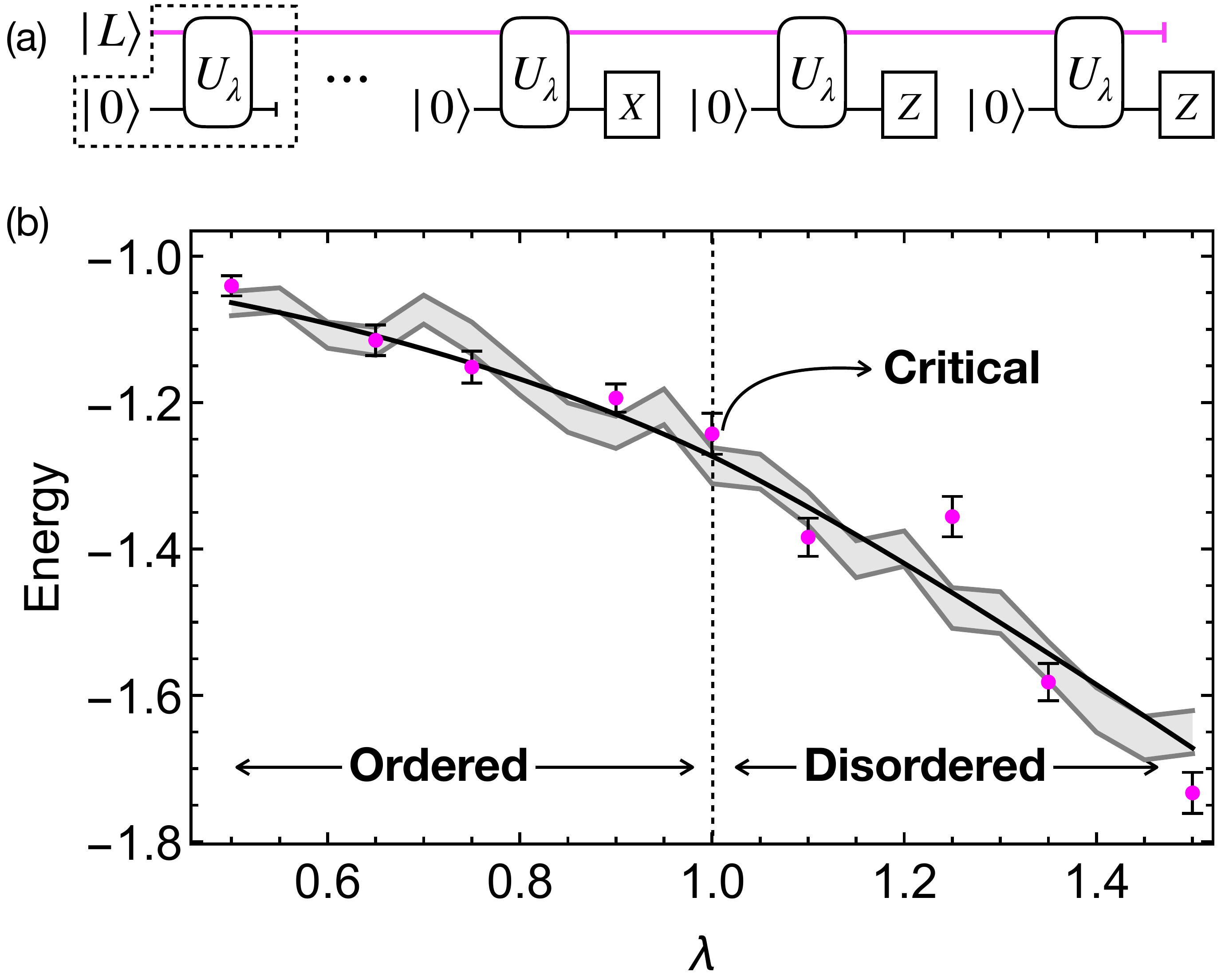}
\caption{(a) Circuits to extract energy estimates from a bond-dimension 2 MPS.  (b) Solid line is the exact ground-state energy of the infinite-size TFIM.  Experimental data (purple, $1\sigma$ error bars) agrees reasonably well from deep inside the ordered phase into the disordered phase.  The gray shaded region is a 1$\sigma$ confidence interval from a numerical simulation with the same shot-count as the experiment.}
\label{fig:energy_plot}
\end{center}
\end{figure}

In order to demonstrate the state preparation method and the extraction of both correlation functions and entanglement entropy, we classically optimize MPS approximations to the TFIM ground state over a range of $\lambda$ and decompose their unitary embeddings into our native gate set. In principle, scaling these techniques to bond dimensions outside the reach of classical MPS optimization should be possible by utilizing parameterized circuits and feeding energy estimates from the quantum computer into a classical optimization routine \cite{Peruzzo:2014aa,foss2020holographic}.  We first perform a scan across the phase transition for a $\chi=2$ MPS, which requires only a single bond qubit.  Since the quantum computer has two gate zones we generally run 2 parallel copies \footnote{Some of the data utilized three parallel implementations per shot, though we found that with only 2 gate zones there was very little benefit in runtime over a $2\times$ parallelization} of the state preparation scheme, making sequential measurements of the system qubit in subsequent bases $X,Z,Z$, as shown in \fref{fig:energy_plot}a. For $j-1$ iterations of the circuit block shown in the black-dashed box in \fref{fig:energy_plot}, this procedure provides sufficient data to estimate both $\langle X_j\rangle$ and $\langle Z_{j+1}Z_{j+2}\rangle$, which (by discrete translational invariance) allows us to reconstruct $\langle H\rangle$ for large enough $j$. In practice, we always choose $j$ sufficiently large that boundary-induced errors are well below anticipated shot noise.  The results of these energy estimates for 5000 total shots per value of $\lambda$ are shown in \fref{fig:energy_plot}(b), and are in good agreement with the exact ground-state energy of the infinite TFIM.

\begin{figure}[!t]
\begin{center}
\includegraphics[width=1.0\columnwidth]{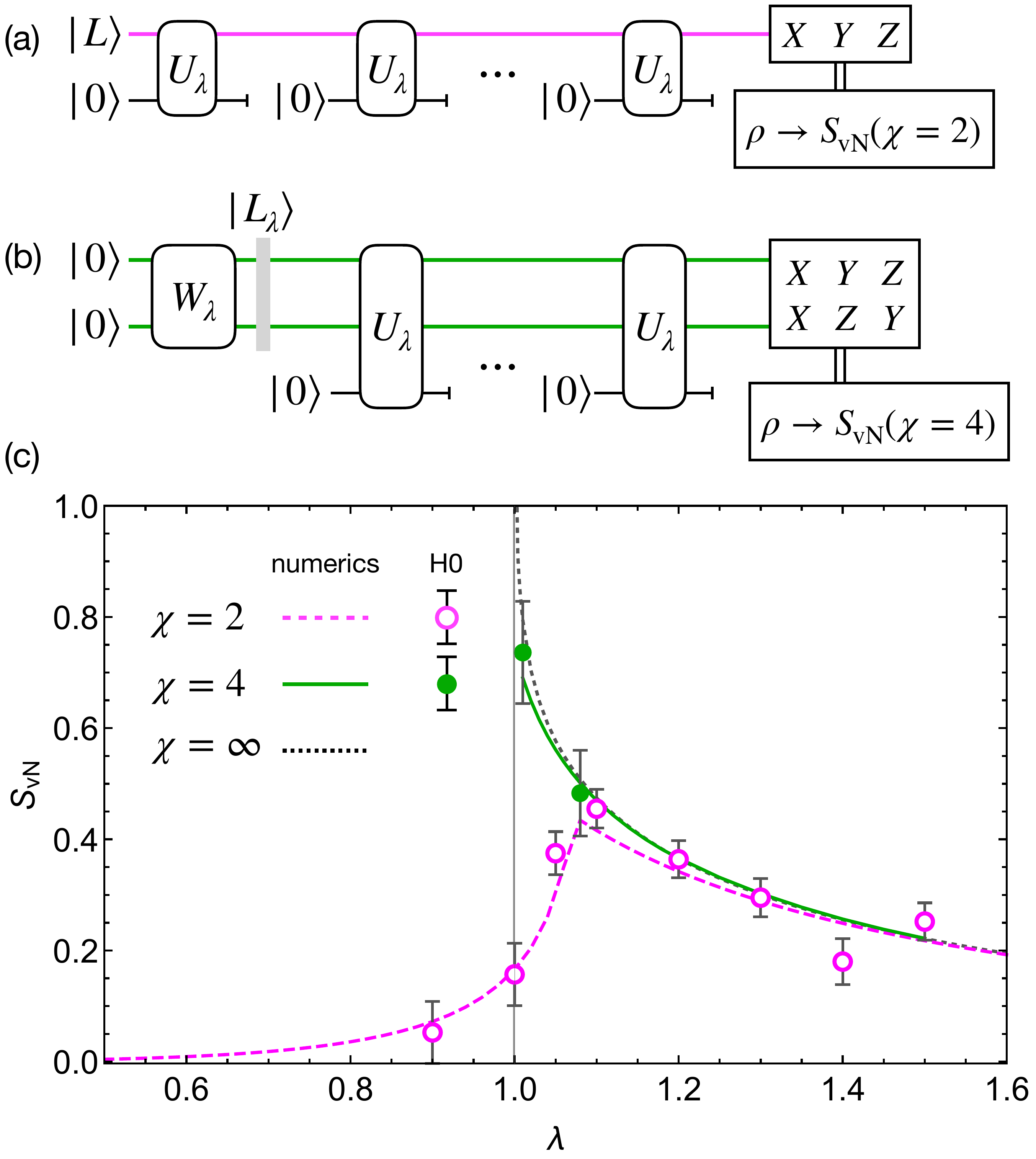}
\caption{(a,b) Circuits for extracting bipartite entanglement entropy of a $\chi=2$ (a) or $\chi=4$ (b) MPS.  (c) Measured entanglement entropy for the TFIM.  Open purple circles are data taken with $n_{b}=1$, and the purple dashed line shows the entanglement entropy of the lowest-energy MPS at $\chi=2^{n_{b}}=2$.  Results using $n_b=2$ are shown as filled green circles, and the solid green line corresponds to the lowest-energy $\chi=4$ MPS.  The black-dotted line is the exact entanglement entropy computed following Ref.\,\cite{PhysRevLett.90.227902}. All error bars represent $\pm\sigma$ confidence intervals obtained by bootstrapping.}
\label{fig:results}
\end{center}
\end{figure}

We extract the TFIM bipartite entanglement entropy by running the same circuits, except instead of making measurements on the system qubit \footnote{In principle, both energy and entanglement entropy measurements could be made in the same circuits, as intermediate measurement of the system qubit does not affect the (marginalized) distribution of bond-qubit measurement outcomes.  However, in order to minimize accumulated cross-talk errors we performed energy extraction and bond-qubit tomography on separate circuits.} we simply extract the bond-register density matrix $\rho_{b}$ by state tomography after the $j^{\rm th}$ iteration.  Results for the entanglement entropy $S_{\rm vN}=-{\rm Tr}(\rho_{b} \log_2\rho_{b})$ are shown as open purple circles in \fref{fig:results}c, and agree well with numerical calculations of the bipartite entanglement entropy for the $\chi=2$ MPS approximation to the ground state.  At any finite bond dimension the peak entropy will also be finite, but the $\chi=2$ peak is also significantly shifted toward the disordered side.  This behavior is expected and reflects the tendency for any ansatz that limits entanglement (e.g., mean-field theory) to overestimate a system's inclination to order.  To better resolve the entropy divergence at the critical point, we took more data just on the disordered side of the transition using one additional bond qubit, giving a $\chi=4$ MPS [filled green circles in \fref{fig:results}(c)], and we see that the growth of entanglement near the true critical point becomes rapidly resolvable upon increasing the size of the bond qubit register.  For comparison, a similarly accurate estimate of the entanglement entropy at the point nearest to the phase transition ($\lambda=1.01$), if achieved by directly preparing the ground state of a large enough system to sufficiently suppress boundary effects, would require $40$ qubits.


Note that all of the data in Figs.\,(\ref{fig:energy_plot},\ref{fig:results}) utilize zero-noise extrapolation \cite{PhysRevLett.119.180509} to mitigate errors on the two-qubit gates (see the supplemental material for details on the error mitigation strategy we employed). For the $\chi=4$ data, we also employ a symmetry-based selection criterion to reduce the number of measurement settings required for tomography of the bond-qubit register.
%
%
To minimize the cost in terms of circuit depth, we let the initial state of the bond-qubit register $\ket{L_{\lambda}}$ be a function of $\lambda$.  Minimization of boundary effects is accomplished by choosing $\ket{L_{\lambda}}$ to have zero overlap onto the eigenstate of the associated MPS channel with smallest non-zero real part of its eigenvalue (this being the most slowly decaying eigenvector), and we find a two-qubit unitary $W_{\lambda}$ to prepare that state, as shown in \fref{fig:results}(b).  Note that this optimization always leaves the bond-qubit in a pure state, so any entropy is due entirely to the application of the MPS channel itself. While the $\chi=4$ results could in principle be continued across the phase transition into the ordered phase, the optimal circuits in the ordered phase are considerably more complex than those in the disordered phase, leading to longer circuit run times and likely invalidating the application of our noise mitigation techniques given current two-qubit gate errors.



\emph{Outlook}---Using a quantum algorithm based on classical tensor-network techniques, we have shown that the entanglement entropy of a formally infinite-size system can be extracted with considerable accuracy and relative ease using a very small number of qubits, giving remarkably good agreement with exact results even in the vicinity of a quantum phase transition.  On a technical level, the algorithm is made possible by recent developments in trapped-ion quantum computing that enable selective mid-circuit measurement and reuse of qubits with extremely low cross-talk errors. Going forward, it would be interesting to combine quantum tensor network algorithms with qubit-efficient schemes to measure R\'enyi entropies \cite{yirka2020qubitefficient} in order to access larger bond-dimension MPS, to extend the present techniques to either tree-like or 2D tensor networks, and to implement quantum analogues of MPS time-evolution algorithms \cite{foss2020holographic}.

\acknowledgements{We thank Karl Mayer for helpful discussions. This work was made possible by a large group of people, and the authors would like to thank the entire Honeywell Quantum Solutions team for their many contributions.   AP was supported by NSF Convergence Accelerator Track C grant OIA 2040549.}

\newpage

\appendix

\section{SUPPLEMENTAL MATERIAL}

\beginsupplement

\subsection{Unitary embedding of MPS for the TFIM}

\begin{figure}[htbp]
\begin{center}
\includegraphics[width=0.95\columnwidth]{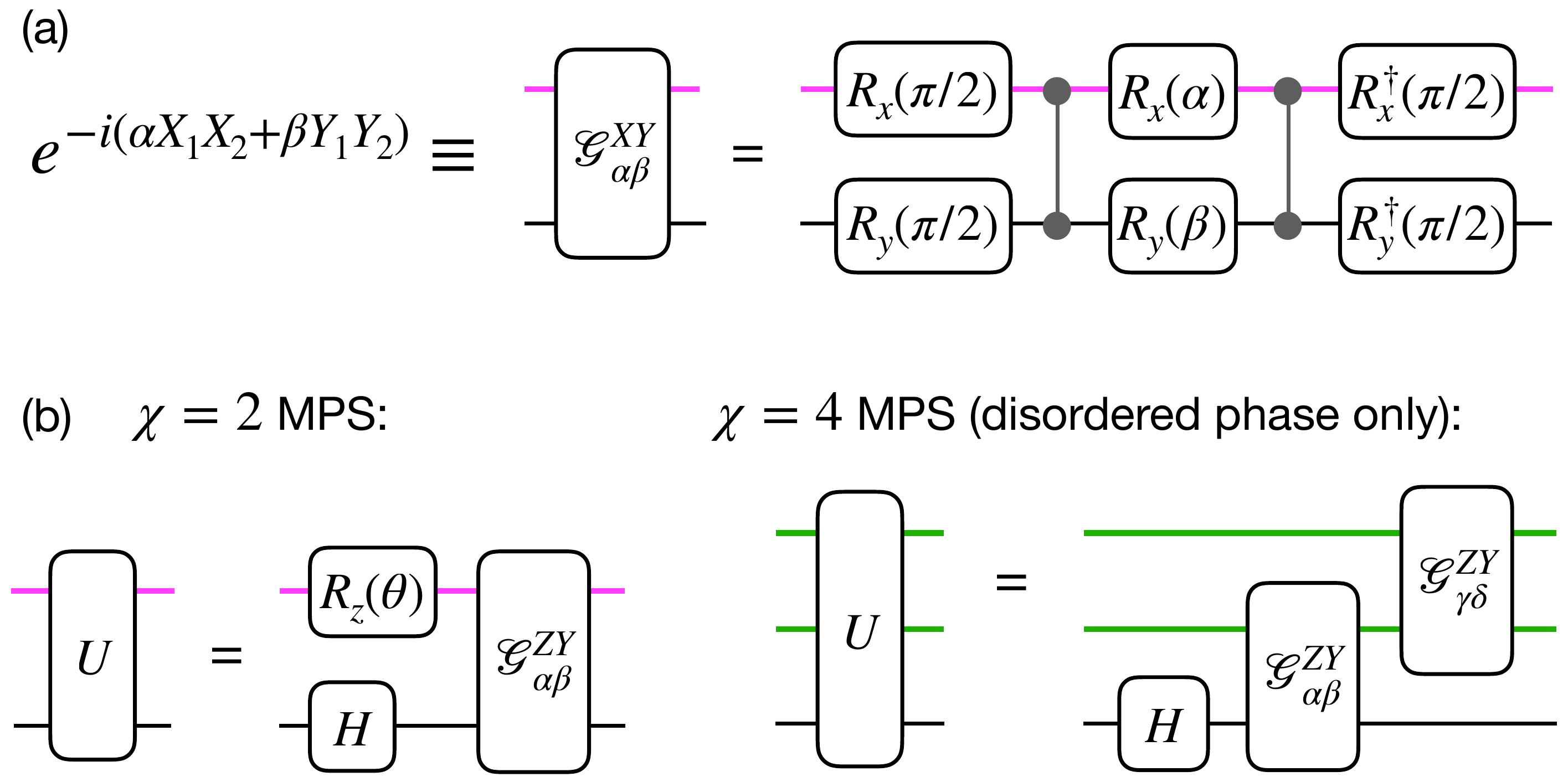}
\caption{(a) The variable angle entangling gate $\mathcal{G}^{XY}_{\alpha\beta}$ is a useful primitive and can be written in terms of $cZ$ gates (which are single-qubit-unitary equivalent to our native entangling gate) as shown. (b) This primitive [or rotations of it, such as $\mathcal{G}^{ZY}=R^{\dagger}_{y}(\pi/2)\mathcal{G}^{XY}R_{y}(\pi/2)$] can be efficiently tiled to produce unitary embeddings of low-bond-dimension MPS for the TFIM.}.
\label{fig:MPS_unitaries}
\end{center}
\end{figure}

For the $\chi=2$ MPS, which has one system qubit and one bond qubit, it is relatively straightforward to parameterize and implement an arbitrary [SU(4)] unitary, into which the isometry $V$ at any $\lambda$ can be embedded.  We use symmetry-based considerations to generate a significantly simplified unitary ansatz, shown on the left-hand side of \fref{fig:MPS_unitaries}(b), which turns out to faithfully embed the isometries of the the lowest-energy $\chi=2$ MPS at all $\lambda$.  For $\chi=4$, we find that tiling this same primitive in a simple fashion [right-hand side of \fref{fig:MPS_unitaries}(b)] captures the optimal MPS approximation to the ground state anywhere in the disordered phase (i.e. on the disordered side of the phase transition \emph{at that bond dimension}).

\subsection{Error mitigation}

In our circuits, the dominant source of error is from two-qubit gates, $U_{zz}\equiv \exp\big[i (\pi/4) Z\otimes Z\big]$, which operate at roughly $99.2\%$ average fidelity.  To mitigate the impact of these errors we perform zero-noise  extrapolation \cite{PhysRevLett.119.180509}: For each circuit $\mathcal{C}_1$, we also run a modified circuit $\mathcal{C}_3$ with the replacement $U_{zz}\rightarrow U_{zz}(U_{zz}^2 [Z\otimes Z])$. In the absence of errors, the term in parentheses is identity and has no impact on the circuit.  If, however, we assume that the single qubit gate $Z$ has no errors and the error on $U_{zz}$ is a depolarizing error channel $\Lambda(p)$ with $p\ll1$, then we have $\Lambda(p)U_{zz}\rightarrow \Lambda(3p) U_{zz}$ (to order $p$).  We can then consider the two probability distributions induced by these circuits, $P_1$ and $P_3$, to belong to a family of distributions $P(p)$ such that $P_1=P(p)$ and $P_3=P(3p)$.  We then linearly extrapolate the error free distribution $P(0)$ from $P_1$ and $P_3$ as $P(0)\approx P(p)-\frac{1}{2}[P(3p)-P(p)]=P_1-\frac{1}{2}(P_3-P_1)$.  We use $P(0)$ to extract approximations of the bond-register density matrix, which we then classically diagonalize to extract the entanglement entropy.

In mapping an MPS to a quantum circuit, the spatial extent of the MPS gets mapped onto the temporal extent of the circuit.  As a result, moving sufficiently far from the boundary into the bulk corresponds to running the circuit (effectively iterating the MPS channel) a sufficiently large number of times, and can lead to fairly deep circuits.  The fact that we are implementing a converging channel with a unique steady state means that, in effect, errors in these circuits tend to heal and only impact observables if they happen (roughly speaking) with in the memory-time of the channel (correlation length of the MPS).  One notable exception to this rule is leakage errors outside the qubit subspace, which amount to a singular perturbation to the MPS channel and completely destroy the steady state.  When leakage errors are small, we expect a quasi-stable steady state to persist over an intermediate time scale, slowly degrading towards a corrupted steady state as leakage errors accrue.

For us, leakage occurs primarily due to spontaneous Raman scattering into the $\ket{S=1/2,F=1,m_{F}=\pm 1}$ Zeeman states during two-qubit gates.  While this contributes a relatively small part ($<0.1\%$) out of the $\sim 0.8\%$ average infidelity of the gate, it can be particularly damaging for the reasons just described. For this reason, we post-select the $\chi=4$ data (which requires significantly more two-qubit gates than the $\chi=2$ data) on data for which the bond-qubits have not leaked by using a leakage-detection gadget similar to Ref.\,\cite{Stricker:2020vx}.

\subsection{Symmetry restricted tomography}

For the $\chi=2$ MPS, which uses only a single bond qubit, we perform full tomography by measuring the bond qubit in the $X$, $Y$, and $Z$ basis.  Defining single qubit Pauli matrices $P_{j=1,\dots,4}\in \{\mathds{1},X,Y,Z\}$, we then estimate the density matrix as
\begin{align}
\rho_{\rm b}\approx \frac{1}{2}\sum_{j}\langle P_j\rangle P_j,
\end{align}
with the expectation values $\langle X\rangle$, $\langle Y\rangle$, and $\langle Z\rangle$ estimated from the experimental data.  For the $\chi=4$ MPS, we can similarly expand the bond-qubit density matrix as
\begin{align}
\rho_{b}=\frac{1}{4}\sum_{jk}P_{j}P_{k}\langle  P_j P_k\rangle,
\end{align}
and the coefficients can always be estimated by taking data in 9 independent measurement settings (choice of $X$, $Y$, or $Z$ on each qubit).  We utilize the observation that the Ising model's $\Z_2$ symmetry (invariance under $Y\!\rightarrow\! Z$, $Z\!\rightarrow \!-Y$) guarantees ${\rm Tr}(P_iP_j)=0$ unless $P_iP_j\in\{ IX,XI,XX,YZ,ZY \}$, which require only 3 measurement settings to extract.

\end{document}